\documentclass[12pt,letterpaper]{revtex4}
\usepackage{hyperref}
\usepackage{makeidx,endnotes,fancyhdr}
\usepackage{showidx}
\usepackage{amsmath,amssymb,theorem,mathrsfs,epsf,eepic}
\usepackage{epsf,epsfig,psfig}
\newcommand{\ket}[1]{|#1\rangle} 
 
 \def\Tr{\operatorname{Tr}}
\def\d{\operatorname{d}} \def\>{\rangle} \def\<{\langle}
\def\sH{\mathscr{H}}  
  
 \def\U1{\mathbb{U}(1)} \def\E{\mathcal{E}}
\def\openone{\mathbb{I}} \def\sL{\mathscr{L}}
\begin{document}

\title{Economical realization of phase covariant devices in arbitrary
  dimensions}
\author{Francesco Buscemi,
Giacomo Mauro D'Ariano, and Chiara Macchiavello} 
\affiliation{QUIT Group, Dipartimento di
  Fisica ``A. Volta'', Universit\`a di Pavia, via
  Bassi 6, I-27100 Pavia, Italy}

\date{\today}

\begin{abstract}
  We describe a unified framework of phase covariant multi user
  quantum transformations for $d$-dimensional quantum systems.  We
  derive the optimal phase covariant cloning and transposition
  tranformations for multi phase states.  We show that for some
  particular relations between the input and output number of copies
  they correspond to economical tranformations, which can be achieved
  without the need of auxiliary systems.  We prove a relation between
  the optimal phase covariant cloning and transposition maps, and
  optimal estimation of multiple phases for equatorial states.
\end{abstract}

\maketitle
\section{Introduction}

The possibility of employing quantum systems with (finite) dimension
higher than two in quantum information has recently triggered much
interest.  In particular, it has been shown that an increase in the
dimension leads to a better performance of various quantum information
protocols, such as for example quantum cryptography~\cite{qc3,3dim} and
some problems in distributed quantum computing~\cite{gima}.  Moreover,
considerable experimental progress has been recently reported in the
generation, manipulation and detection of quantum systems with higher
dimension~\cite{exp}.

In this work we consider the case of phase covariant transformations
where the information is encoded into phase properties of states with
arbitrary finite dimension $d$. Encoding information into phase shifts
has important applications in quantum computation and quantum
information. For example, it was shown that the existing quantum
algorithms can be described in a unified way as quantum interference
processes among different computational paths where the result of the
computation is retrieved from a phase shift~\cite{cemm}.

We will describe in a unified framework the features of multi users
phase covariant transformations in arbitrary dimension $d$, where
typically an arbitrary number of input systems $N$ described by the
same quantum state is transformed into a larger number of output
systems $M$, which are still described by the same output density
operators.  We will then specify this description to two tasks of
interest in quantum information, namely cloning and phase conjugation.

The no cloning theorem~\cite{no-cloning} states the impossibility of
perfectly cloning unknown quantum states selected from a nonorthogonal
set, and is the basis of the security of quantum
cryptography~\cite{BB84,E91}. Approximate quantum cloning has been
extensively studied in the last years and has led to relevant results
in quantum cryptography. The eavesdropping strategies in quantum key
distribution protocols that are known to be optimal so far are
actually based on cloning attacks~\cite{3dim,bcdm,cbm}.  Moreover,
quantum cloning allows to study the sharing of quantum information
among several parties and it may be applied also to study the security
of multi-party cryptographic schemes~\cite{multi-qc}.

Perfect phase conjugation of the density operator of an unknown input
state, or equivalently ideal time reversal, is also forbidden by the
laws of quantum mechanics.  Such a transformation has also recently
attracted much interest in connection to the problem of entanglement,
in regards to the so-called PPT (positive partial transpose)
criterion~\cite{peres,H3}. Since phase conjugation cannot be achieved
unitarily, one can try to approximate the transformation with a
physical channel, optimizing the fidelity of the output state with the
complex-conjugated input. In the case of qubits ($d=2$) phase
conjugation is unitarily equivalent to the NOT
operation~\cite{univ-not}. For the set of all pure states the
resulting universal optimal channel is
``classical''~\cite{univ-not,opt-transp}, namely, it can be achieved
by state-estimation followed by state-preparation. In contrast, as we
will show in this work, for equatorial pure states the optimal phase
covariant conjugation map is a purely quantum transformation (for any
number of input copies), generalising to the case of many copies the
analogous result already proved for a single input
system~\cite{phasenot}.

The paper is organised as follows. In Sect. \ref{s:pcg} we describe in
a unified framework the operation of multi users phase covariant
transformations in arbitrary finite dimension $d$.  In Sect.
\ref{s:ec} we review the concept of economical maps.  In Sects.
\ref{sec:cloning} and \ref{sec:not} we derive the optimal phase
covariant cloning and phase conjugation maps respectively for
equatorial input states, and show that for some particular relations
between the input and output number of copies the optimal
transformations can be achieved economically.  In Sect.
\ref{sec:estimation} we prove a relation between optimal multiple
phase estimation procedures and the optimal cloning and phase
conjugation maps.  In Sect. \ref{s:conc} we summarise the results
presented in this work.  Some technical details of the derivations
presented in the paper are explained in the Appendices.

\section{Phase covariant devices}
\label{s:pcg}

In this paper we consider quantum devices, or \emph{channels} (i.~e.
trace-preserving completely positive maps~\cite{kraus}), from states on
an input quantum system $\sH_\textrm{in}$ to states on a generally
different output quantum system $\sH_\textrm{out}$, for which we
assume that an underlying global symmetry under the action of the
phase rotations group $\U1$ holds. More precisely, we will optimise
the action of such devices on pure $d$-dimensional input states of the
form
\begin{equation}
  |\psi(\{\phi_j\})\>=\frac{1}{\sqrt d}(\ket{0}+e^{i\phi_1}\ket{1}
  +e^{i\phi_2}\ket{2}+\cdots+e^{i\phi_{d-1}}\ket{d-1}),\label{qudits}
\end{equation}
where $\{|0\>,\dots,|d-1\>\}$ is a fixed orthonormal basis of a
$d$-dimensional system $\sH$, and the $\phi_j$'s are $(d-1)$
independent phases in the interval $[0,2\pi)$. Notice that the choice
$\phi_0=0$ is not restrictive, since an overall phase is negligible.
In the case of qubits, i.~e. $d=2$, pure states as in
Eq.~(\ref{qudits}) all lie on one equator of the Bloch sphere and they
clearly form a set that is invariant under rotations around the fixed
axis orthogonal to this equator. These rotations form a group that is
isomorphic to the group $\U1$.  For generic dimensions $d>2$, this
geometrical picture is straightforwardly generalized by saying that
pure states in Eq.~(\ref{qudits}) form a set of states that is
invariant under the action of the unitary representation
$U_{\{\phi_j\}}=|0\>\<0|+\sum_{j=1}^{d-1}|j\>\<j|e^{i\phi_j}$ of the
group $\U1^{\times(d-1)}$. In the following, with a little abuse of
terminology, we will call states of the form (\ref{qudits}) as
equatorial states, also in the case $d>2$.  Notice that starting from
a fixed state $|\psi_0\>=d^{-1/2}\sum_i|i\>$, usually called
\emph{seed}, it is possible to span the whole invariant family by
applying the unitary operator $U_{\{\phi_j\}}$
\begin{equation}
U_{\{\phi_j\}}|\psi_0\>=|\psi(\{\phi_j\})\rangle.
\end{equation}

Since we are considering input states belonging to
phase-rotations-group-invariant families, the natural scenario for our
analysis is then the framework of \emph{phase covariant channels},
namely channels $\E$ that automatically propagate to the output the
action of the group on the input as follows
\begin{equation}\label{eq:covariance}
  \E(V_g\rho V_g^\dag)=W_g\E(\rho)W_g^\dag,
\end{equation}
where $V_g$ and $W_g$ are unitary representations of
$\U1^{\times(d-1)}$ on the input and output space, respectively. More
explicitly, when the input consists of $N$ copies of an unknown pure
equatorial state, i.~e.  $|\psi(\{\phi_j\})\>^{\otimes N}$, we have
$V_g=V_{\{\phi_j\}}=U_{\{\phi_j\}}^{\otimes N}$. The choice of the
output representation $W_g$ will depend on the task we want to
optimize. For the moment, just notice that both $V_g$ and $W_g$ are
\emph{different} unitary representations of the \emph{same} group
$\U1^{\times(d-1)}$.

Since we are considering only pure input states of the form
$|\psi\>^{\otimes N}$ we can restrict our attention to channels whose
input states have support on the symmetric subspace $\sH^{\otimes
  N}_+$ of $\sH^{\otimes N}$, that is, $\sH_\textrm{in}=\sH^{\otimes
  N}_+$. Moreover, we require that also the output states have support
on the symmetric subspace, namely $\sH_\textrm{out}=\sH^{\otimes
  M}_+\subset\sH^{\otimes M}$. In this way it is guaranteed that the
output single site density operators are the same. For the following,
we choose an orthonormal basis in the symmetric subspace $\sH^{\otimes
  N}_+$ of the form
\begin{equation}\label{symmetrization}
  |\{n_i\}\>_N=\ket{n_0,n_1,n_2,\dots,n_{d-1}}_N=\frac{1}{\sqrt{N!}}\sum_{\{\pi\}}P_\pi^{(N)}|\underbrace{00\dots0}_{n_0}\underbrace{11\dots1}_{n_1}\dots\underbrace{d-1\dots d-1}_{n_{d-1}}\>,
\end{equation}
where $P_\pi^{(N)}$ denotes a permutation operator of the $N$
systems, $n_0$ is the number of systems in state $\ket{0}$, $n_1$ in
state $\ket{1}$, and so on, with the constraint $\sum_{i=0}^{d-1}
n_i=N$. The notation $|\{m_i\}\>_M$, with $\sum_{i=0}^{d-1}m_i=M$,
denotes the analogous symmetric state as in Eq.~(\ref{symmetrization})
for the output subspace $\sH^{\otimes M}_+$. As a convention, in this 
paper we will consistently use letters $n$'s for the input and $m$'s for the
output.

A convenient formalism to deal with covariant channels is the
Choi-Jamio\l{}kowski isomorphism~\cite{jam,choi} between
\emph{completely positive} maps $\mathcal{M}$ from states on
$\sH_\textrm{in}$ to states on $\sH_\textrm{out}$ and \emph{positive}
operators $R_\mathcal{M}$ on $\sH_\textrm{out}\otimes\sH_\textrm{in}$
\begin{equation}\label{eq:choi-jam}
  \mathcal{M}\longleftrightarrow R_\mathcal{M}=(\mathcal{M}\otimes\mathcal{I})|\Omega\>\<\Omega|
\end{equation}
where $\mathcal{I}$ is the identity channel and
$|\Omega\>=\sum_{k=1}^d|k\>\otimes|k\>$ is the (non normalized)
maximally entangled vector in the Hilbert space
$\sH_\textrm{in}\otimes\sH_\textrm{in}$. With the notation introduced
in Eq.~(\ref{symmetrization}) we have
\begin{equation}\label{eq:maxent}
  |\Omega\>=\sum_{\{n_i\}}|\{n_i\}\>_N\otimes|\{n_i\}\>_N.
\end{equation}
The correspondence~(\ref{eq:choi-jam}) is
one-to-one, the inverse formula being
\begin{equation}\label{eq:reconstruction}
  \mathcal{M}(\rho)=\Tr_\textrm{in}[(\openone_\textrm{out}\otimes\rho^*)\ R_\mathcal{M}],
\end{equation}
where $\Tr_\textrm{in}$ denotes the trace over $\sH_\textrm{in}$,
$\openone_\textrm{out}$ is the identity matrix over
$\sH_\textrm{out}$, and $\rho^*$ is the complex conjugated of $\rho$
with respect to the basis fixed by $|\Omega\>$ in
Eq.~(\ref{eq:maxent}). Trace-preservation condition is then given by
$\Tr_\textrm{out}[R_\mathcal{M}]=\openone_\textrm{in}$.

In terms of the Choi-Jamio\l{}kowski operator, the covariance
condition~(\ref{eq:covariance}) can be rewritten as a commutation
relation~\cite{paolino}
\begin{equation}
[R_\mathcal{E},W_g\otimes V_g^*]=0,
\label{eq:commutation}
\end{equation}
where $W_g\otimes V_g^*$ is a new unitary representation of
$\U1^{\times(d-1)}$. Such a representation is generally reducible,
whence, by Schur lemma, $R_\mathcal{E}$ splits into a direct sum
\begin{equation}
R_\mathcal{E}=\bigoplus_\alpha R_\mathcal{E}^\alpha,
\end{equation}
where the index $\alpha$ labels the equivalence classes of the
one-dimensional~\cite{onedim} irreducible representations of
$W_g\otimes V_g^*$. In Sections~\ref{sec:cloning} and~\ref{sec:not} we
will specialize Eq.~(\ref{eq:commutation}) to the cases of $N\to M$
cloning and $N\to M$ phase conjugation, for which
$W_g=U_{\{\phi_j\}}^{\otimes M}$ and
$W_g=\left(U_{\{\phi_j\}}^*\right)^{\otimes M}$, respectively.

\section{Economical maps}
\label{s:ec}
Let $\mathcal{M}$ be a completely positive, trace-preserving map from
states on $\sH_\textrm{in}$ to states on $\sH_\textrm{out}$. From the
Stinespring representation theorem~\cite{Stinespring}, it immediately
follows that for every completely positive trace-preserving map it is
possible to find an auxiliary quantum system with Hilbert space $\sL$
and an isometry $V$ from $\sH_\textrm{in}$ to
$\sH_\textrm{out}\otimes\sL$, $V^\dag V=\openone_\textrm{in}$, such
that
\begin{equation}\label{Stine}
\mathcal{M}(\rho)=\Tr_\sL[V\rho V^\dag].
\end{equation}
Starting from Eq.~(\ref{Stine}), it is always possible to construct a
unitary interaction $U$ realizing $\mathcal{M}$~\cite{ozawa,sc} as
follows
\begin{equation}\label{unit-real}
\mathcal{M}(\rho)=\Tr_\sL\left[U(\rho\otimes|a\>\<a|)U^\dag\right],
\end{equation}
where $|a\>$ is a fixed pure state of a \emph{second} auxiliary
quantum system, say $\sL' $, such that $\sH_\textrm{in}\otimes\sL'=
\sH_\textrm{out}\otimes\sL $. The Hilbert spaces $\sL$ and $\sL'$ are
generally different, and actually play different physical roles.

We define a trace-preserving completely positive map $\mathcal{M}$ to
be \emph{economical} if and only if it admits a unitary form $U$ as
\begin{equation}\label{econ-real}
\mathcal{M}(\rho)=U(\rho\otimes|a\>\<a|)U^\dag,
\end{equation}
namely, if and only if the map can be physically realized without
discarding---i.~e. tracing out---resources. We can simply prove that
the only maps admitting an economical unitary implementation $U$ as in
Eq.  (\ref{econ-real}) are those for which
\begin{equation}\label{isometrical}
\mathcal{M}(\rho)=V\rho V^\dag
\end{equation}
for an isometry $V$, $V^\dag V=\openone_\textrm{in}$. In fact,
since
$(\openone_\textrm{in}\otimes\<a|)U^\dag
U(\openone_\textrm{in}\otimes|a\>)=\openone_\textrm{in}$,
$U(\openone_\textrm{in}\otimes|a\>)$ is an isometry from
$\sH_\textrm{in}$ to $\sH_\textrm{out}\otimes\sL$. On the other
hand, starting from Eq.~(\ref{isometrical}) and using the Gram-Schmidt method
one can extend any
isometry $V$ from $\sH_\textrm{in}$ to $\sH_\textrm{out}\otimes\sL$ to
a unitary $U$ on the same output space, and write it in the form
$V=U(\openone_\textrm{in}\otimes|a\>)$ for a unit vector $|a\>\in\sL'$,
with $\sH_\textrm{in}\otimes\sL'=\sH_\textrm{out}\otimes\sL $.

Considering classical resources as free, the most general definition
of economical map corresponds to having a \emph{random-unitary}
realization of the form
\begin{equation}
\mathcal{M}(\rho)=\sum_ip_iU_i(\rho\otimes|a\>\<a|)U^\dag_i,
\end{equation}
where $p_i\ge 0$, $\sum_ip_i=1$. Using the same fixed ancilla state
$|a\>$ for all indeces $i$ is not a loss of generality, since in
constructing the operators $U_i$'s there is always freedom in the
choice of the vector $|a\>$. According to this more general
definition, all economical maps can always be written as a
randomization of the form (\ref{isometrical}) as follows
\begin{equation}
\mathcal{M}(\rho)=\sum_ip_iV_i\rho V^\dag_i.
\end{equation}

\section{Phase covariant cloning}
\label{sec:cloning}

In this section we derive the form of quantum channels $\mathcal{C}$ 
that best approximate the ideal cloning map
\begin{equation}
  |\psi(\{\phi_j\})\>^{\otimes N}\longmapsto|\psi(\{\phi_j\})\>^{\otimes M},
\label{eq:impossible-clon}\end{equation}
with $M>N$, for all possible values $\phi_j\in[0,2\pi)$. In this case 
the choice of the unitary representation on the output space
is clearly $W_g=U_{\{\phi_j\}}^{\otimes M}$.  The
commutation relation~(\ref{eq:commutation}) can then be rewritten as
\begin{equation}
[R_\mathcal{C},U_{\{\phi_j\}}^{\otimes M}\otimes
  (U^*_{\{\phi_j\}})^{\otimes N}]=0\;.
\label{eq:cloning-comm}\end{equation}
>From the above equation it follows that $R_\mathcal{C}$ splits into
the block-form
\begin{equation}\label{block-form}
R_\mathcal{C}=\bigoplus_{\{m_j\}}R_{\{m_j\}},
\end{equation}
where each set of values $\{m_j\}$ identifies a unique class of
equivalent irreducible representations of $U_{\{\phi_j\}}^{\otimes
  M}\otimes(U^*_{\{\phi_j\}})^{\otimes N}$. The equivalent
representations within each class can be conveniently written, using
the symmetrization convention as in Eq.~(\ref{symmetrization}), as
\begin{equation}
  \left\{\ket{m_0+n_0,m_1+n_1,m_2+n_2,\dots,m_{d-1}+n_{d-1}}_M\otimes\ket{n_0,n_1,n_2,\dots,n_{d-1}}_N\right\}_{\{n_i\}},
\end{equation}
with $\sum_{i=0}^{d-1}n_i=N$ and $\sum_{j=0}^{d-1}m_j=M-N$. The
multi-index $\{n_i\}$ runs over all orthonormal vectors of the
symmetrised basis for $\sH^{\otimes N}_+$. With this notation,
Eq.~(\ref{block-form}) becomes
\begin{equation}
R_\mathcal{C}=\sum_{\{m_j\}}\sum_{\{n'_i\},\{n''_i\}}r^{\{m_j\}}_{\{n'_i\},\{n''_i\}}|\{m_j\}+\{n'_i\}\>\<\{m_j\}+\{n''_i\}|_M\otimes|\{n'_i\}\>\<\{n''_i\}|_N.
\end{equation}

We now have to adjust the parameters
$\left\{r^{\{m_j\}}_{\{n'_i\},\{n''_i\}}\right\}$ describing a generic
channel satisfying the commutation relation~(\ref{eq:cloning-comm}),
in order to shape $R_\mathcal{C}$ to optimally approximate the ideal
map~(\ref{eq:impossible-clon}). Such an optimal approximation
reasonably maximizes the fidelity $F_\mathcal{C}$ between the ideal
output, namely $|\psi(\{\phi_j\})\>^{\otimes M}$, and the actual
channel output
$\mathcal{C}(|\psi(\{\phi_j\})\>\<\psi(\{\phi_j\})|^{\otimes N})$. By
exploiting the inverse formula~(\ref{eq:reconstruction}) and the
commutation relation~(\ref{eq:cloning-comm}), one has
\begin{equation}
  F_\mathcal{C}=\Tr[|\psi_0\>\<\psi_0|^{\otimes(N+M)}R_\mathcal{C}].
\end{equation}
Another commonly adopted figure of merit 
is the single-site fidelity $F_\mathcal{C}^1$ between the
ideal output $|\psi(\{\phi_j\})\>$ and the actual single-site output
$\Tr_{M-1}[\mathcal{C}(|\psi(\{\phi_j\})\>\<\psi(\{\phi_j\})|^{\otimes
  N})]$, namely
\begin{equation}
F_\mathcal{C}^1=\Tr[|\psi_0\>\<\psi_0|\otimes\openone^{\otimes(M-1)}\otimes|\psi_0\>\<\psi_0|^{\otimes N}R_\mathcal{C}].
\end{equation}
We point out that a channel $\mathcal{C}$ maximizing $F_\mathcal{C}$ does not
necessarily maximize also $F_\mathcal{C}^1$~\cite{keyl-werner,nostro}.

When the output number of copies takes the form
$M=kd+N$, with $k\in\mathbb{N}$ and $d$ the dimension of $\sH$, there
exists a unique channel maximizing at the same time both
$F_\mathcal{C}$ and $F_\mathcal{C}^1$~\cite{pheconclon}. 
Such a channel is described by the positive rank-one operator
\begin{equation}\label{eq:RC}
  R_\mathcal{C}=|r_{\{k\}}\>\<r_{\{k\}}|,
\end{equation}
where
\begin{equation}\label{optmap2}
  |r_{\{k\}}\>=\sum_{\{n_j\}}|k+n_0,\dots,k+n_i,\dots\>_M\otimes
  |n_0,\dots,n_i\dots\>_N,\qquad\sum_jn_j=N.
\end{equation}
The corresponding single-site fidelity $F_\mathcal{C}^1$ takes the
form
\begin{equation}\label{eq:FNM}
  F_\mathcal{C}^1=\frac{1}{d}+\frac{1}{Md^{N+1}}\sum_{\{\bar{n}_j\}}\sum_{i\neq j}\frac{N!}{\bar{n}_0!\dots\bar{n}_i!\dots\bar{n}_j!\dots}\sqrt{\frac{(\bar{n}_i+k+1)(\bar{n}_j+k+1)}{(\bar{n}_i+1)(\bar{n}_j+1)}},\qquad M=kd+N,
\end{equation}
where, for the sake of symmetry of the formula, we have chosen the
multi index $\bar{n}_j$ such that $\sum_j\bar{n}_j=N-1$.  In the case
$N=1$, the above equation is simplified as
\begin{equation}\label{eq:F1M}
  F_\mathcal{C}^1=\frac{1}{d}+\frac{(d-1)(M+d-1)}{Md^2},\qquad M=kd+1,
\end{equation}
since $\sum_{i\neq j}(k+1)=kd(d-1)+d(d-1)=(d-1)(M+d-1)$. Notice that
$F_\mathcal{C}^1$ is always strictly greater than the analogous
optimal fidelity for the universal cloner~\cite{keyl-werner}, that is
$F^1_\textrm{univ}=\frac{2M+d-1}{M(d+1)}$. This is due to the fact
that we are now imposing a covariance condition under the action of
$\U1^{\times(d-1)}$, that is a much looser condition~\cite{subgroup}
than imposing covariance under the action of the whole universal group
$\mathbb{SU}(d)$, and therefore there is more freedom in adjusting
free parameters to obtain better performances.

As a final remark notice that, since $R_\mathcal{C}$ is rank-one, the
channel $\mathcal{C}$ acts as (this can be simply checked by using the
inverse formula~(\ref{eq:reconstruction}))
\begin{equation}
\mathcal{C}(\rho^{\otimes N})=V\rho^{\otimes N}V^\dag,
\end{equation}
where $V$ is the isometry defined as
\begin{equation}
  V|n_0,n_1,\dots,n_i,\dots\>_N=|n_0+k,n_1+k,\dots,n_i+k,\dots\>_M.
\end{equation}
According to the definitions of Section~\ref{s:ec}, this implies that
$\mathcal{C}$ is an economical map, and therefore it does not require
additional resources other than the $(M-N)$ input blank copies in
order to be unitarily realized. This is in contrast to what happens in
the universal case, for which $M$ additional sytems must be
provided~\cite{opt-transp,cerf} besides the $N$ input copies.

\section{Phase conjugation}
\label{sec:not}

Another basic device which is impossible to achieve in the framework
of quantum mechanics is the NOT gate, where the Bloch vector of any
input states is reversed, or equivalently the phase conjugation
operation.  In this section we will derive the form of the quantum
channels $\mathcal{N}$ that optimally approximate the operation of
phase conjugation
\begin{equation}
  |\psi(\{\phi_j\})\>^{\otimes N}\longmapsto(|\psi(\{\phi_j\})\>^*)^{\otimes M}=|\psi(\{-\phi_j\})\>^{\otimes M},
\end{equation}
with $M>N$, for all possible values $\phi_j\in[0,2\pi)$. The case
$M=N=1$ has been thoroughly analysed~\cite{phasenot}.
In the case of phase conjugation the output
unitary representation of $\U1^{\times(d-1)}$ must be chosen as
$W_g=(U^*_{\{\phi_j\}})^{\otimes M}$ and the commutation
relation~(\ref{eq:commutation}) becomes
\begin{equation}
[R_\mathcal{N}, (U^*_{\{\phi_j\}})^{\otimes(M+N)}]=0.
\label{eq:not-comm}
\end{equation}
As in the case of phase covariant cloning, the above relation implies
a decomposition of $R_\mathcal{N}$ into the block-form
$R_\mathcal{N}=\bigoplus_{\{m_j\}}R_{\{m_j\}}$, where each set of
values $\{m_j\}$ identifies a unique class of equivalent irreducible
representations of $(U^*_{\{\phi_j\}})^{\otimes(M+N)}$. The equivalent
representations within each class can be conveniently written as
\begin{equation}
  \left\{\ket{m_0-n_0,m_1-n_1,m_2-n_2,\dots,m_{d-1}-n_{d-1}}_M\otimes\ket{n_0,n_1,n_2,\dots,n_{d-1}}_N\right\}_{\{n_i\}},
\end{equation}
with $\sum_{i=0}^{d-1}n_i=N$ and $\sum_{j=0}^{d-1}m_j=M+N$. It is
clear that the previous equation is well defined only when $m_i\ge
n_i$, for all $i$. In the following we will see that, when the
analytical optimization is possible, such a condition is always
satisfied.

The figure of merit that we will consider to approximate the phase conjugation 
channel is the single-site
fidelity
\begin{equation}
  F_\mathcal{N}^1=\Tr[|\psi_0\>\<\psi_0|\otimes\openone^{\otimes(M-1)}\otimes|\psi_0\>\<\psi_0|^{\otimes N}R_\mathcal{N}].
\end{equation}
where
\begin{equation}
R_\mathcal{N}=\sum_{\{m_j\}}\sum_{\{n'_i\},\{n''_i\}}r^{\{m_j\}}_{\{n'_i\},\{n''_i\}}|\{m_j\}-\{n'_i\}\>\<\{m_j\}-\{n''_i\}|_M\otimes|\{n'_i\}\>\<\{n''_i\}|_N.
\end{equation}

Exploiting  cumbersome combinatorial calculations very similar
to those reported in previous work~\cite{pheconclon}, it can be
shown that, in the case $M=kd-N$, with $k\in\mathbb{N}\ge N$ and $d$
the dimension of $\sH$, there exists a unique channel maximizing
$F_\mathcal{C}^1$.  Such a channel is described by the positive rank-one
operator 
\begin{equation}
  R_\mathcal{N}=|r_{\{k\}}\>\<r_{\{k\}}|,
\end{equation}
where
\begin{equation}
  |r_{\{k\}}\>=\sum_{\{n_j\}}|k-n_0,\dots,k-n_i,\dots\>_M\otimes
  |n_0,\dots,n_i\dots\>_N,\qquad\sum_jn_j=N,
\end{equation}
and it acts as an isometrical embedding
\begin{equation}
\mathcal{N}(\rho^{\otimes
  N})=V\rho^{\otimes N}V^\dag,
\end{equation}
where the isometry $V$ is defined as
\begin{equation}
  V|n_0,n_1,\dots,n_i,\dots\>_N=|k-n_0,k-n_1,\dots,k-n_i,\dots\>_M.
\end{equation}
Therefore, also  the optimal phase conjugation map for output number of copies 
$M=kd-N$ can therefore be realised economically. 
Its single-site fidelity $F_\mathcal{N}^1$ is given by
\begin{equation}\label{eq:FnotNM}
  F_\mathcal{N}^1=\frac{1}{d}+\frac{1}{Md^{N+1}}\sum_{\{\bar{n}_j\}}\sum_{i\neq j}\frac{N!}{\bar{n}_0!\dots\bar{n}_i!\dots\bar{n}_j!\dots}\sqrt{\frac{(k-\bar{n}_i)(k-\bar{n}_j)}{(\bar{n}_i+1)(\bar{n}_j+1)}},\qquad M=kd-N,
\end{equation}
where $\sum_j\bar{n}_j=N-1$.
Since $\sum_{i\neq j}k=kd(d-1)=(d-1)(M+1)$
in the case $N=1$ the above expression is simplified as
\begin{equation}\label{eq:Fnot1M}
  F_\mathcal{N}^1=\frac{1}{d}+\frac{(d-1)(M+1)}{Md^2}\qquad M=kd-1.
\end{equation}
Notice that for qubits $F_\mathcal{N}^1=F_\mathcal{C}^1$. 
This is due to the fact that for equatorial qubits perfect 
phase conjugation can be achieved unitarily by a $\pi$ rotation along the 
$x$ axis. Optimal phase conjugation therefore is equivalent to optimal 
phase covariant cloning followed by such a rotation, which does not decrease 
the cloning fidelity. In all the other cases with $d>2$, 
$F_\mathcal{N}^1$ is always strictly
smaller than $F_\mathcal{C}^1$. Actually, in these cases phase conjugation 
can be performed only approximately and therefore the global transformation 
corresponding to a generation of many phase conjugated copies is worse than 
just cloning them.
However, in the limit of large number of output copies, i.~e.
$M\to\infty$, they both tend to the same limit, as we will show in 
Section~\ref{sec:estimation}.

\section{Relations with optimal phase  estimation}
\label{sec:estimation}

\begin{figure}[htb]
\epsfig{file=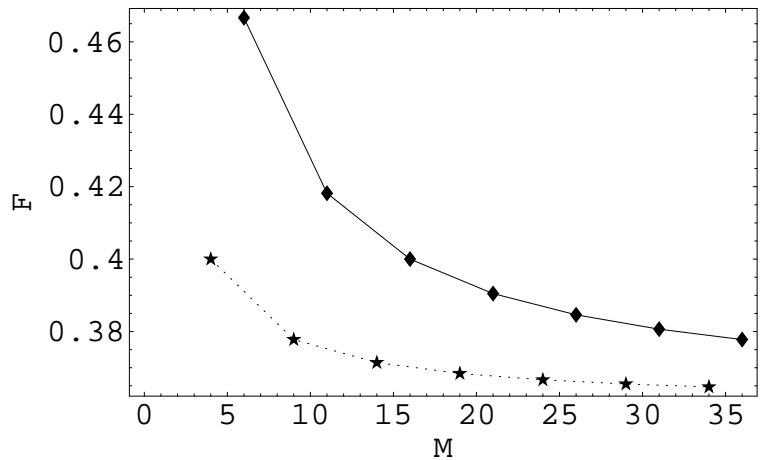,width=10cm}
\caption{Comparison between single-site fidelities of phase covariant
  $1\to M$ optimal cloning (continuous line) and phase conjugation
  (dotted line), for $d=5$. Both curves tend to the limit of
  9/25=0.36, that is, the fidelity of optimal phase estimation.}
\label{fig:1}
\end{figure}

Both the cloning fidelity $F_\mathcal{C}^1$ in Eq.~(\ref{eq:FNM}) and
the phase conjugation fidelity $F_\mathcal{N}^1$ in
Eq.~(\ref{eq:FnotNM}), in the limit $M\to\infty$, that is,
$k\to\infty$ with $M\approx kd$, take the form
\begin{equation}\label{FPEN}
  F^1=\frac{1}{d}+\frac{1}{d^{N+2}}\sum_{\{\bar{n}_i\}}\sum_{i\neq j}\frac{N!}{\bar{n}_0!\dots}\frac{1}{\sqrt{(\bar{n}_i+1)(\bar{n}_j+1)}},\qquad\sum_i\bar{n}_i=N-1.
\end{equation}
The above expression coincides with the single-site fidelity
$F_\mathcal{P}^1$ of optimal phase estimation on $N$ copies of
equatorial states~\cite{multi-phase}. For all possible values of $N$
and $M$, the following relations then hold
\begin{equation}\label{eq:limits}
  F_\mathcal{C}^1\ge F_\mathcal{N}^1\ge F_\mathcal{P}^1,\qquad\lim_{M\to\infty}F_\mathcal{C}^1=\lim_{M\to\infty}F_\mathcal{N}^1=F_\mathcal{P}^1.
\end{equation}
The above inequalities are illustrated in Figure~\ref{fig:1}, where
the optimal fidelities of phase covariant cloning and phase
conjugation are reported for equatorial states with $d=5$ and $N=1$.
First of all, let us notice that phase covariant conjugation,
contrarily to the case of universal transposition~\cite{opt-transp}
for which it is known that the optimal strategy trivially consists of
an estimation followed by a suitable preparation, achieves a fidelity
$F_\mathcal{N}^1$ that is always greater than the fidelity
$F_\mathcal{P}^1$ achievable by means of a measure-and-prepare scheme.
Moreover, Eq.~(\ref{eq:limits}) confirms the general fact that cloning
fidelity, in the limit of infinite output copies number, tends to
state estimation fidelity, and shows that this also holds for other
symmetrical covariant devices, such as phase conjugation.

Here we prove that not only the fidelities $F_\mathcal{C}^1$ and
$F_\mathcal{N}^1$ tend to the phase estimation fidelity
$F_\mathcal{P}^1$, but also that optimal phase covariant cloning
$\mathcal{C}$ and phase conjugation $\mathcal{N}$ maps tend, in the
limit, to the phase estimation map $\mathcal{P}$ (which estimates the
phases $\{\overline{\phi_j}\}$ and reprepares the state
$|\psi(\{\overline{\phi_j}\})\>$). This is clearly a much stronger
statement than that concerning just fidelities~\cite{note-on-estim}.
The main ingredient we need for the proof is that the single-site
output state coming from the channel $\mathcal{C}$ can be parametrised
by a shrinking parameter $\eta_{\mathcal{C}}$ as
\begin{equation}\label{eq:reduced-clon}
  \Tr_{M-1}\left[\mathcal{C}\left(|\psi(\{\phi_j\})\>\<\psi(\{\phi_j\}|\right)^{\otimes N})\right]=\eta_{\mathcal{C}}|\psi(\{\phi_j\})\>\<\psi(\{\phi_j\}|+(1-\eta_{\mathcal{C}})\frac{\openone}{d},
\end{equation}
with $\eta_{\mathcal{C}}=(d F_{\mathcal{C}}^1-1)/(d-1)$. Analogous
formulas hold for phase conjugation $\mathcal{N}$ and phase estimation
$\mathcal{P}$, as a consequence of the phase covariant property of the
maps (for the explicit calculations, see
Appendices~\ref{sec:appendixa} and~\ref{sec:appendixb}).

The proof then goes through a concatenation argument. Imagine to
perform an optimal phase estimation~\cite{multi-phase} over $N$ copies
of the unknown state $|\psi(\{\phi_j\})\>$. After obtaining the
optimal estimated value $\{\overline{\phi_j}\}$ of the phases, it is
possible to prepare $M$ copies of the state
$|\psi(\{\overline{\phi_j}\})\>$.  This procedure is, by definition, a
sub-optimal phase covariant $N\to M$ cloning: the fidelity of such $M$
copies must be smaller than (or at most equal to) the fidelity of the
output of an optimal phase covariant cloner, that is
\begin{equation}\label{est-le-clon}
  F_\mathcal{P}^1(N)\le F_\mathcal{C}^1(N,M),\qquad\forall N,M.
\end{equation}
(We put in parentheses the dependence of the fidelities on the input
number of copies $N$ and the output number $M$).

The opposite direction can be proved by concatenating the optimal
$N\to M$ phase covariant cloner with the optimal \emph{state}
estimation described in References~\cite{DBE,BM}. Since a state
estimation implies a phase estimation, it is possible to interpret the
whole procedure as a sub-optimal phase estimation: the single-site
fidelity $\overline{F}^1(N,M)$ obtained in this sub-optimal way must
be smaller than or equal to the optimal phase estimation fidelity
$F_\mathcal{P}^1(N)$ for all possible values of $M$. The state
estimation map $\mathcal{S}$ works as follows~\cite{BM}
\begin{equation}
\mathcal{S}(\rho^{\otimes M})=\eta_\mathcal{S}\rho+(1-\eta_\mathcal{S})\frac{\openone}{d}.
\end{equation}
Applying $\mathcal{S}$ to the output of the phase covariant $N\to M$
cloner, we get
\begin{equation}
  \mathcal{S}(\mathcal{C}(|\,\psi(\{\phi_j\})\,\>\<\,\psi(\{\phi_j\})\,
|^{\otimes N}))=\eta_\mathcal{S}\Tr_{M-1}\left[\mathcal{C}(|\,\psi(\{\phi_j\})
\,\>\<\,\psi(\{\phi_j\})\,|^{\otimes N})\right]+(1-\eta_\mathcal{S})
\frac{\openone}{d}\;.
\end{equation}
Since we assumed that the output state $\rho_{M}$ of a phase covariant
cloner has support on the symmetric subspace it can be linearly
decomposed as $\rho_{M}= \sum_i\lambda_i|\psi_i\>\<\psi_i|^{\otimes
  M}$ with $\sum_i\lambda_i=1$~\cite{paolo-vittorio}. Therefore the
above expression can be written as
\begin{equation}
  \mathcal{S}(\mathcal{C}(|\,\psi(\{\phi_j\})\,\>\<\,\psi(\{\phi_j\})\,
|^{\otimes N}))=\eta_\mathcal{S}\eta_\mathcal{C}|\,\psi(\{\phi_j\})\,\>\<\,\psi(\{\phi_j\})\,|+(\eta_\mathcal{S}(1-\eta_\mathcal{C})+(1-\eta_\mathcal{S}))\frac{\openone}{d}.
\end{equation}
Noticing that the shrinking factor
$\eta_\mathcal{S}=M/(M+d)$~\cite{BM} approaches unit for $M\to\infty$,
the last equation implies that
\begin{equation}
  \lim_{M\to\infty}\overline{F}(N,M)=\lim_{M\to\infty}F_\mathcal{C}(N,M),
\end{equation}
and according to the previous remark about the sub-optimality of this
phase estimation procedure, one has
\begin{equation}\label{clon-le-est}
  \lim_{M\to\infty}F_\mathcal{C}(N,M)\le F_\mathcal{P}(N),\qquad\forall N.
\end{equation}
Eq.~(\ref{clon-le-est}) together with Eq.~(\ref{est-le-clon}) prove
Eq.~(\ref{eq:limits}).

The above argument can be applied also to the case of phase
conjugation. Actually, a suboptimal phase covariant conjugation map
can be achieved by first performing an optimal phase estimation on the
input equatorial states, which gives the estimated values
$\{\overline{\phi_j}\}$ for the phases, and then preparing $M$ copies
of the state $|\psi(\{-\overline{\phi_j}\})\>$. Moreover, a suboptimal
phase estimation can be realised by first applying an optimal $N\to M$
phase covariant conjugation device and then performing optimal state
estimation on the $M$ output states.  In this way we would be able to
estimate the $(d-1)$ phase values of the $M$ output states and we
would have an estimate of the phases of the $N$ input states just by
changing the signs. The comparison of the above two procedures allows
to establish the equivalence of optimal phase estimation and optimal
phase covariant transposition in the limit of infinite number of
output copies.

\section{Conclusions}
\label{s:conc}

In this paper we have studied the efficiency of phase covariant multi
user channels in arbitrary finite dimension.  In particular, we have
derived the form of the channels that optimally approach quantum
cloning and phase conjugation for multi-phase equatorial states.  We
have shown that for certain relations between the input and output
number of copies the optimal transformations can be achieved
economically.  We have derived a relation between the above mentioned
transformations and optimal multiple phase estimation procedures.  In
the case of phase conjugation we have shown that, in contrast to the
customary case of the Universal-NOT on qubits (or the universal
conjugation in arbitrary dimension), the optimal phase covariant
transformation for equatorial states is a nonclassical channel, which
cannot be achieved via a measurement/preparation procedure.

\appendix

\section{Single-site reduced output state of optimal phase estimation}\label{sec:appendixa}

The phase estimation channel $\mathcal P$ working over $N$ copies of
the input state $|\psi(\{\phi_j\})\>$ can be regarded as a machine
preparing the state $|\psi(\{\overline{\phi_j}\})\>$ according to the
estimated phases values $\{\overline{\phi_j}\}$. The output state
$|\psi(\{\overline{\phi_j}\})\>$ is prepared with probability density
\begin{equation}\label{prob-density}
  p(\{\overline{\phi_j}\})
  =\left|\<\psi(\{\phi_j\})^{\otimes N}|e(\{\overline{\phi_j}\})\>\right|^2,
\end{equation}
where
\begin{equation}
  |e(\{\overline{\phi_j}\})\>=U_{\{\overline\phi_j\}}^{\otimes N}\sum_{\{n_i\}}|\{n_i\}\>_N,\qquad\sum_in_i=N,
\end{equation}
namely, a generalized Susskind-Glogower state~\cite{sussglo}, and
$|e(\{\overline{\phi_j}\})\>\<e(\{\overline{\phi_j}\})|$ is the POVM
density of the optimal (multi)-phase estimation~\cite{multi-phase}
over $N$ copies. Using the formalism of quantum operations, the
single-site reduced output state of such apparatus can be simply
written as
\begin{equation}\label{ph-est-map}
  \mathcal{P}(|\psi(\{\phi_j\})\>\<\psi(\{\phi_j\})|^{\otimes N})=\int\frac{\d\{\overline{\phi_j}\}}{(2\pi)^{d-1}}p(\{\overline{\phi_j}\})|\psi(\{\overline{\phi_j}\})\>\<\psi(\{\overline{\phi_j}\})|.
\end{equation}

By covariance, we can exploit the calculations only for the input
$|\psi_0\>$ and then generalize trivially to all possible input states
$|\psi(\{\phi_j\})\>$ considered here. From Eqs.~(\ref{prob-density})
and~(\ref{ph-est-map}), the starting point is
\begin{equation}\label{starting}
\begin{split}
  \mathcal{P}(|\psi_0\>\<\psi_0|^{\otimes N})&=\int\frac{\d\{\overline{\phi_j}\}}{(2\pi)^{d-1}}\Tr\left[|\psi_0\>\<\psi_0|^{\otimes N}|e(\{\overline{\phi_j}\})\>\<e(\{\overline{\phi_j}\})|\right]|\psi(\{\overline{\phi_j}\})\>\<\psi(\{\overline{\phi_j}\})|\\
  &=\Tr_N\left[\openone\otimes|\psi_0\>\<\psi_0|^{\otimes
      N}\int\frac{\d\{\overline{\phi_j}\}}{(2\pi)^{d-1}}|\psi(\{\overline{\phi_j}\})\>\<\psi(\{\overline{\phi_j}\})|\otimes|e(\{\overline{\phi_j}\})\>\<e(\{\overline{\phi_j}\})|\right].
\end{split}
\end{equation}
Recalling the orthogonality relation
\begin{equation}
\int\frac{\d\gamma}{2\pi}\exp[i(m-n)\gamma]=\delta_{mn},\qquad\forall m,n\in\mathbb{Z},
\end{equation}
and the explicit expression for
$|e(\{\overline{\phi_j}\})\>\<e(\{\overline{\phi_j}\})|$
\begin{equation}
  |e(\{\overline{\phi_j}\})\>\<e(\{\overline{\phi_j}\})|=U(\{\overline{\phi_j}\})^{\otimes N}\left[\sum_{\{n'_i\},\{n''_j\}}|\{n'_i\}\>\<\{n''_j\}|\right]U^\dag(\{\overline{\phi_j}\})^{\otimes N},
\end{equation}
we have
\begin{equation}\label{orthogonality}
\begin{split}
\int&\frac{\d\{\overline{\phi_j}\}}{(2\pi)^{d-1}}|\psi(\{\overline{\phi_j}\})\>\<\psi(\{\overline{\phi_j}\})|\otimes|e(\{\overline{\phi_j}\})\>\<e(\{\overline{\phi_j}\})|\\
&=\sum_{\{n_i\}}\sum_{i,j}\frac{|i\>\<j|}{d}\otimes|\{n_i\}\>\<n_0,\dots,n_i-1,\dots,n_j+1,\dots|.
\end{split}
\end{equation}
Substituting Eq.~(\ref{orthogonality}) into Eq.~(\ref{starting}), we
get the formula we were looking for, namely
\begin{equation}
  \mathcal{P}(|\psi_0\>\<\psi_0|)=\frac{\openone}{d}+\frac{1}{d^{N+1}}\sum_{\{\bar{n}_i\}}\sum_{i\neq j}\frac{N!}{\bar{n}_0!\dots}\frac{1}{\sqrt{(\bar{n}_i+1)(\bar{n}_j+1)}}|i\>\<j|,\qquad\sum_j\bar{n}_j=N-1,
\end{equation}
whence the single-site fidelity~(\ref{FPEN}) of (multi)-phase
estimation.

\section{Single-site reduced output state of optimal phase covariant
  cloning and optimal phase conjugation}\label{sec:appendixb}

Here we explicitly derive the general form of the reduced output state
of the phase-covariant $N\to M$ cloner in Eq.~(\ref{eq:reduced-clon}).
(The phase conjugation case is completely analogous.) From
Eqs.~(\ref{eq:reconstruction}) and~(\ref{eq:RC}):
\begin{equation}
\begin{split}
  \Tr&_{M-1}[\mathcal{C}(|\psi_0\>\<\psi_0|^{\otimes N})]\\
  &=\Tr_{M-1,N}\left[\openone^{\otimes M}\otimes|\psi_0\>\<\psi_0|^{\otimes N}\;\sum_{\{n'_i\},\{n''_j\}}|n'_0+k,\dots\>\<n''_0+k,\dots|_M\otimes|n'_0,\dots\>\<n''_0,\dots|_N\right]\\
  &=\frac{1}{d^N}\sum_{\{n'_i\},\{n''_j\}}\left[\binom{N}{n'_0;n'_1;\dots}\binom{N}{n''_0;n''_1;\dots}\right]^{1/2}\Tr_{M-1}\left[|n'_0+k,\dots\>\<n''_0+k,\dots|_M\right]\\
  &=\frac{1}{d^N}\sum_{\{n'_i\},\{n''_j\}}\left[\binom{N}{n'_0;n'_1;\dots}\binom{N}{n''_0;n''_1;\dots}\right]^{1/2}\left[\binom{M}{n'_0+k;\dots}\binom{M}{n''_0+k;\dots}\right]^{-1/2}\\
  &\phantom{0000000}\times\Tr_{M-1}\left[|\widetilde{n'_0+k,\dots}\>\<\widetilde{n''_0+k,\dots}|_M\right]\\
  &=T_\textrm{diag}+T_\textrm{off-diag},
\end{split}
\end{equation}
where
\begin{equation}
|\widetilde{n_0+k,\dots}\>_M=\sum_{\{\pi\}}P_\pi^{(M)}|\underbrace{00\dots0}_{n_0+k}\underbrace{11\dots1}_{n_1+k}\dots\underbrace{d-1\dots d-1}_{n_{d-1}+k}\>
\end{equation}
is a non normalized vector, with the notation of
Eq.~(\ref{symmetrization}). In order to make the calculation clearer,
we split the previous equation in its diagonal part:
\begin{equation}
\begin{split}
T_\textrm{diag}&=\frac{1}{d^N}\sum_{\{n'_i\},\{n''_j\}}\sum_i\left[\binom{N}{n'_0;n'_1;\dots}\binom{N}{n''_0;n''_1;\dots}\right]^{1/2}\left[\binom{M}{n'_0+k;\dots}\binom{M}{n''_0+k;\dots}\right]^{-1/2}\\
&\phantom{000000000000}\times\Tr_{M-1}\Big[|i\>\<i|\otimes|n'_0+k,\dots,n'_i+k-1,\dots\>\<n''_0+k,\dots,n''_i+k-1,\dots|\Big]\\
&\phantom{000000000000}\times\left[\binom{M-1}{n'_0+k;\dots;n'_i+k-1;\dots}\binom{M-1}{n''_0+k;\dots;n''_i+k-1;\dots}\right]^{1/2}\\
&=\frac{1}{Md^N}\sum_{\{n_i\}}\frac{N!}{n_0!n_1!\dots}\sum_i(n_i+k)|i\>\<i|,
\end{split}
\end{equation}
and its off-diagonal part:
\begin{equation}
\begin{split}
T_\textrm{off-diag}&=\frac{1}{Md^N}\sum_{\{n_i\}}\sum_{i\neq j}\frac{N!}{n_0!\dots (n_i-1)!\dots n_j!\dots}\sqrt{\frac{(n_i+k)(n_j+k+1)}{n_i(n_j+1)}}|i\>\<j|\\
&=\frac{1}{Md^N}\sum_{\{\bar{n}_i\}}\sum_{i\neq j}\frac{N!}{\bar{n}_0!\dots \bar{n}_i!\dots \bar{n}_j!\dots}\sqrt{\frac{(\bar{n}_i+k+1)(\bar{n}_j+k+1)}{(\bar{n}_i+1)(\bar{n}_j+1)}}|i\>\<j|,
\end{split}
\end{equation}
with the constraints $\sum_j n_j=N$ and $\sum_j\bar{n}_j=N-1$.

First of all, notice that the reduced state is correctly normalized
since $\sum_{\{n_i\}} N!/(n_0!\dots)=d^N$ and $\sum_i(n_i+k)=M$, and
that the fidelity with respect to $|\psi_0\>$ is precisely
$F_\mathcal{C}^1(N,M)$ in Eq.~(\ref{eq:FNM}), since
\begin{equation}
\Tr\left[|\psi_0\>\<\psi_0|\sum_i\frac{n_i+k}{M}|i\>\<i|\right]=\frac{1}{d}
\end{equation}
and
\begin{equation}
\Tr\left[|\psi_0\>\<\psi_0|\sum_{i\neq j}\sqrt{\frac{(\bar{n}_i+k+1)(\bar{n}_j+k+1)}{(\bar{n}_i+1)(\bar{n}_j+1)}}|i\>\<j|\right]=\frac{1}{d}\sum_{i\neq j}\sqrt{\frac{(\bar{n}_i+k+1)(\bar{n}_j+k+1)}{(\bar{n}_i+1)(\bar{n}_j+1)}}.
\end{equation}
Moreover, looking at the expressions of $T_\textrm{diag}$ and
$T_\textrm{off-diag}$ involving a sum over \emph{all} possible
multi-indeces $\{n_i\}$, one can recognize that the diagonal entries
are all multiplied by the same coefficient, as well as the
off-diagonal ones. The reduced output state can then be written as
\begin{equation}
\Tr_{M-1}[\mathcal{C}(|\psi_0\>\<\psi_0|^{\otimes N})]=\eta_\mathcal{C}(N,M)|\psi_0\>\<\psi_0|+(1-\eta_\mathcal{C}(N,M))\frac{\openone}{d}.
\end{equation}

\section*{Acknowledgments}

This work has been supported in part by EC under the project SECOQC 
(Contract No. IST-2003-506813) and by the Italian MIUR under PRIN 2005.


\begin{thebibliography}{99}
\bibitem{onedim} The group $\U1^{\times(d-1)}$ is commutative, whence
  its irreducible representations are all one-dimensional.
\bibitem{subgroup} Actually, $\U1^{\times(d-1)}$ is a proper subgroup
  of $\mathbb{SU}(d)$.
\bibitem{note-on-estim} Recently, it has been proved that cloning
  channels, in the limit of infinite output copies, tend to
  measure-and-prepare channels~\cite{acin}. Here, we are able to
  explicitly show how fast this limit is reached, for every finite
  $M$.
\bibitem{paolo-vittorio} This holds by linearity, since every
  symmetric operator $O$ can be written as a linear combination of
  $N$-fold tensor product pure states, namely
  $O=\sum_i\lambda_i|\psi_i\>\<\psi_i|^{\otimes N}$. See
  Reference~\cite{symmetric}.
\end{thebibliography}
\end{document}